\documentclass[10pt,twocolumn,letterpaper]{article}

\usepackage{iccv}              

\definecolor{iccvblue}{rgb}{0.21,0.49,0.74}
\usepackage[pagebackref,breaklinks,colorlinks,allcolors=iccvblue]{hyperref}
\usepackage{colortbl}
\usepackage{comment}
\def\paperID{9150} 
\def\confName{ICCV}
\def\confYear{2025}

\title{Gain-MLP: Improving HDR Gain Map Encoding via a Lightweight MLP}

\author{Trevor D. Canham$^1$ \hspace{2em} SaiKiran Tedla$^1$ \hspace{2em} Michael J. Murdoch$^2$ \hspace{2em} Michael S. Brown$^1$\\
$^1$York University \hspace{2em}
$^2$Rochester Institute of Technology}

\begin{document}
\maketitle

\newcommand{\sai}[1]{{\textbf{\textcolor{magenta}{[SAI] }}\textcolor{magenta}{#1}}}

\begin{abstract}
While most images shared on the web and social media platforms are encoded in standard dynamic range (SDR), many displays now can accommodate high dynamic range (HDR) content. Additionally, modern cameras can capture images in an HDR format but convert them to SDR to ensure maximum compatibility with existing workflows and legacy displays. To support both SDR and HDR, new encoding formats are emerging that store additional metadata in SDR images in the form of a gain map. When applied to the SDR image, the gain map recovers the HDR version of the image as needed. These gain maps, however, are typically down-sampled and encoded using standard image compression, such as JPEG and HEIC, which can result in unwanted artifacts. In this paper, we propose to use a lightweight multi-layer perceptron (MLP) network to encode the gain map. The MLP is optimized using the SDR image information as input and provides superior performance in terms of HDR reconstruction. Moreover, the MLP-based approach uses a fixed memory footprint (10 KB) and requires no additional adjustments to accommodate different image sizes or encoding parameters.
We conduct extensive experiments on various MLP based HDR embedding strategies and demonstrate that our approach outperforms the current state-of-the-art.
\end{abstract}    
\section{Introduction}
\label{sec:intro}

\begin{figure}[]
\centering
\includegraphics[width=0.48\textwidth]{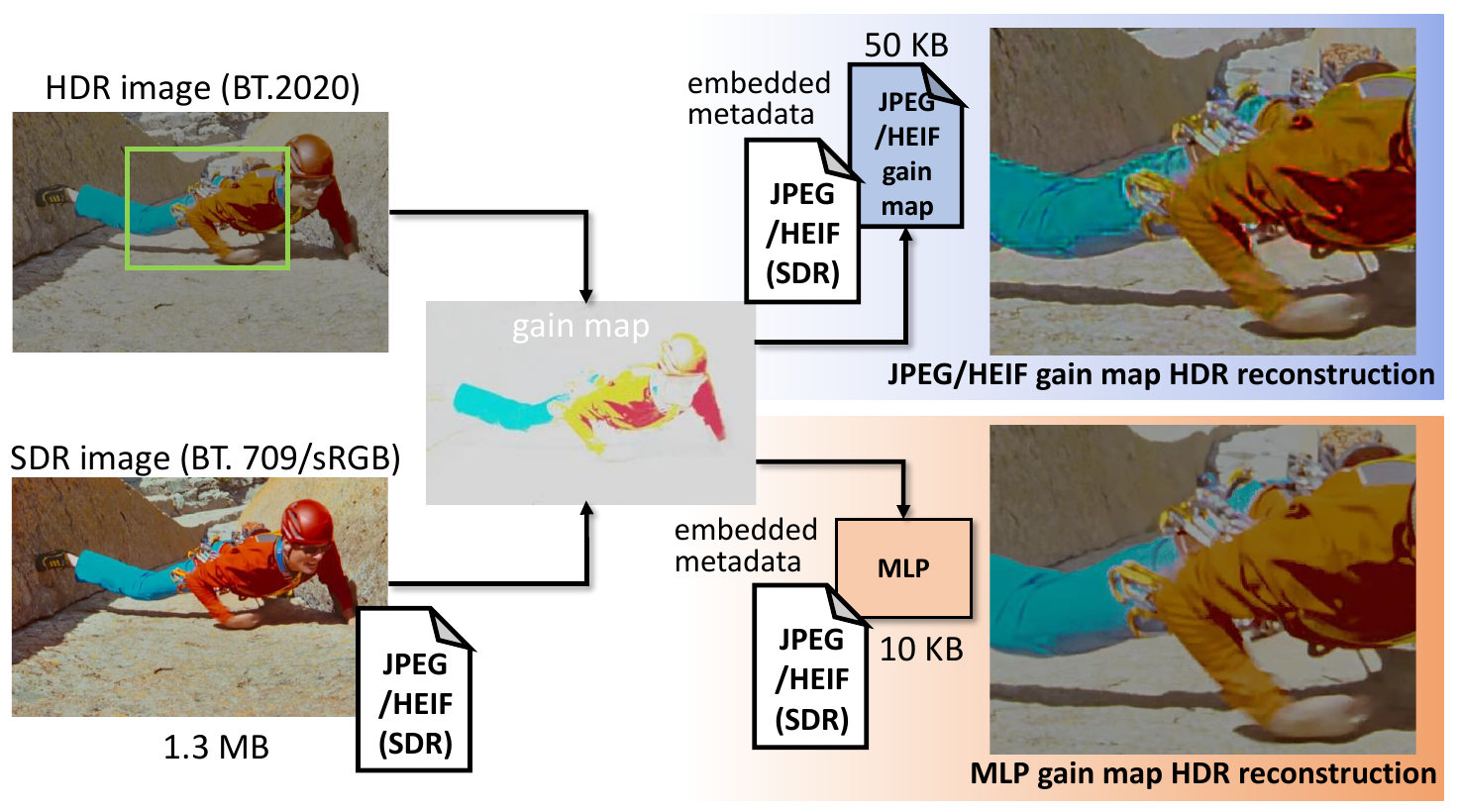}
\caption{High-dynamic-range (HDR) reconstruction methods embed HDR metadata in a standard dynamic range (SDR) image file in the form of a per-channel gain map. Current approaches rely on 8-bit JPEG (or HEIF) to encode the gain map. We show that a lightweight MLP (10 KB) that uses the SDR image as input provides better reconstruction with minimum memory overhead.}\vspace{-6mm}
\label{fig:teaser}
\end{figure}

High dynamic range (HDR) and wide color gamut displays are now commonly found in smartphones and tablets. HDR displays enable contrast ratios that mimic real-world scenes with the ability to produce high-quality luminance gradations from $0.01$ to $1000$ cd/m$^2$ (nits), unlike the standard dynamic range (SDR) displays, which are limited to $1$ to $100$ cd/m$^2$~\cite{reinhard20}. 
Additionally, cameras---particularly smartphone cameras---can now natively capture images in HDR format. However, to ensure compatibility with existing SDR displays, most images are still encoded in SDR formats, such as the BT.709 \cite{bt709} (sRGB) standard used in 8-bit JPEG.

\vspace{10mm}

The availability of better-quality displays has created a need for new, flexible encoding strategies that accommodate both legacy SDR devices and newer HDR displays. Major device manufacturers, software companies, and standards organizations are actively developing encoding recommendations for adaptive dynamic range formats. Examples include Apple's Extended Dynamic Range (EDR), Android's UltraHDR, Samsung's SuperHDR, Adobe's gain map specification, and the ISO 21496 recommendation for digital photography~\cite{wd21496}, which allow for images to be appropriately adapted to varying brightness settings in real-time.

These strategies all agree on storing pixel-wise gain maps that allow easy translation or interpolation between image versions intended for SDR and HDR display. Specifically, gain maps, calculated as a ratio between SDR and HDR pixel values, are quantized to 8-bits and compressed in a multiple image format (e.g., HEIF, JPEG-XL, etc.) to serve as metadata for the primary 8-bit SDR image. This gain map can be decompressed and used to reconstruct the HDR version of the image by multiplying it with the SDR image. This reliance on conventional compression, however, can lead to unwanted artifacts even at standard compression quality factors, while increasing the compression quality leads to increased memory size.  

\paragraph{\textbf{Contribution}}~To address the need for high-quality gain map encoding with minimal data overhead, we demonstrate how recent implicit neural representations can outperform traditional encoding techniques such as JPEG, HEIC, and JPEG-XL. Specifically, we introduce the novel application of multi-layer perceptron (MLP) networks utilizing color and positional coordinates $(r,g,b,x,y)$ to achieve an exceptionally lightweight encoding of gain maps (10 KB) with minimal reconstruction artifacts. Furthermore, through extensive experimentation, we show that using an exponential residual encoding of the gain map enables the MLP network to outperform state-of-the-art methods across a wide range of bit rates. As part of this work, we extend an existing manual tone mapping dataset with alternative canonical tone mappers, allowing for a more thorough characterization of this applied problem space. Finally, we present a novel use of chromatic noise patterns, which exhibit characteristics of natural image statistics, for initializing implicit neural representation networks.
\section{Related Work}
\label{sec:related}

Related work is discussed for areas: (1) tone mapping, which is required to produce the SDR and HDR encoding; (2) inverse tone mapping; and (3) MLP-based neural implicit functions.

\begin{figure*}[t]
\centering
\includegraphics[width=1\textwidth]{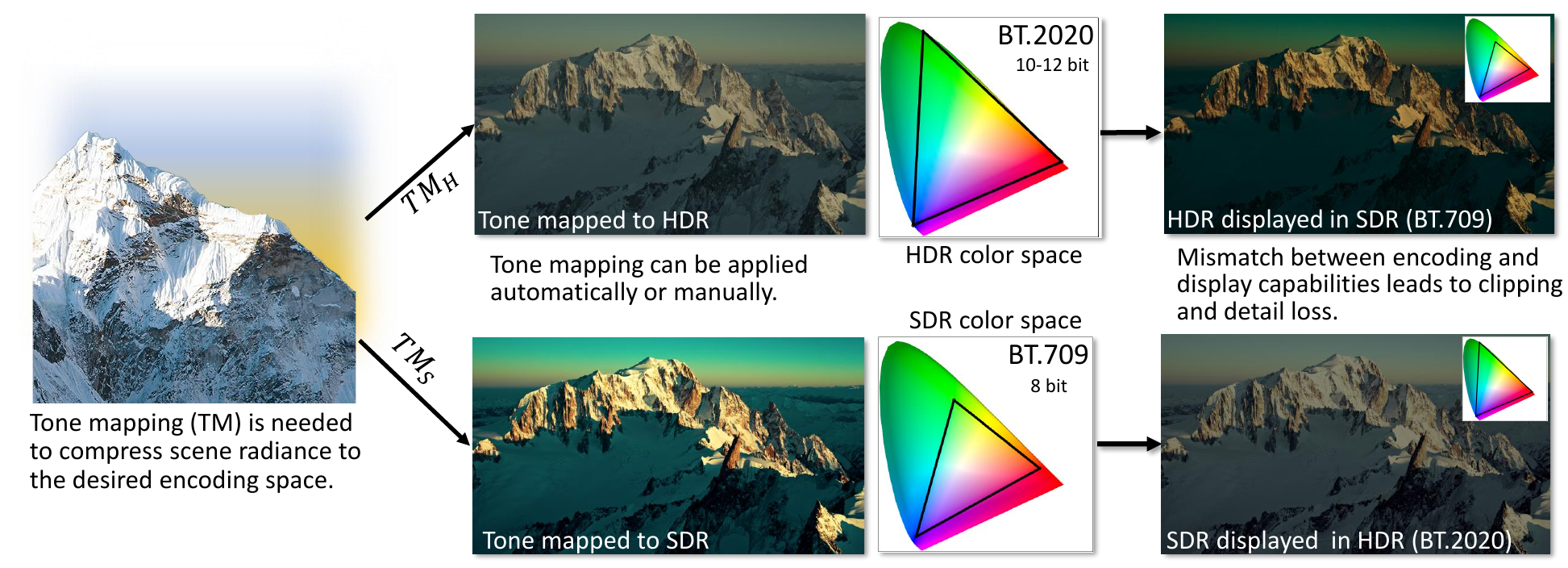}
\caption{Scene radiance values must be compressed (i.e., tone-mapped) to fit a desired target range for display.  This can be accomplished via tone mapping algorithms or manually by colorists. The common SDR image format, BT.709~\cite{bt709} (i.e., sRGB), is a small gamut color space that can be sufficiently encoded using 8-bits per channel.   The widely used HDR format, BT.2020 ~\cite{bt2020} is a wide gamut color space that recommends 10--12 bit encoding.  This figure shows the relative size difference of these encodes in chromaticity space.  Clipping and detail loss can occur when the encoding does not match the display's capabilities (e.g., displaying an HDR image on an SDR display or vice versa). An effective means to properly convert between the two tone mapped encodings is needed.  Note that this PDF is intended for viewing on an SDR/sRGB display. As a result, the HDR images shown in this paper will appear desaturated.} 
\label{fig:tm}
\end{figure*}

\subsection{Tone Mapping}
An open problem since the advent of photography, tone mapping involves redistributing the luminance values of real scenes to render convincing images in the limited luminance range of printed, projected, or displayed images~\cite{giorgianni09}.
Fig. \ref{fig:tm} shows an example where $TM_H$ and $TM_S$ represent tone mapping to HDR and SDR, respectively.  Tone mapping is inherently a one-to-many problem as SDR display encoding necessitates enhancing some scene details at the expense of others.   As a result, existing tone mapping methods achieve varying computational efficiency and qualitative advantages, often favoring certain scene types over others (e.g., indoor, landscape, portrait), but the ideal display rendering remains in the realm of artistic interpretation. 

In the traditional approaches to this problem \cite{tumblin93,reinhard02}, all pixels are processed through the same non-linear mapping function, which is derived as a function of the global dynamic ranges of the source scene and target display. Later advancements included local operators \cite{reinhard05}.
Ou et al. \cite{ou22} provide a recent review of tone mapping methods adapted for real-time hardware implementation, which details the dominant strategies employed in modern camera pipelines.  While the gain map framework is agnostic to the tone mapping operator used to generate the SDR and HDR encoding, the choice of tone mapping algorithm directly influences the gain map, which can in turn impact compression \cite{mai13}. 

In the experiments of Sec. \ref{sec:exp}, we choose a representative method from three categories of tone mapping methods: exposure/filtering-based \cite{reinhard05}, log-based \cite{drago03}, histogram equalization based \cite{larson97}. However, photographers and cinema colorists often prefer to accomplish the task manually. This lack of consensus is explained by the one-to-many nature of the problem and the difficulty in subjective evaluation of the results \cite{ashikhmin06}.

\subsection{Inverse Tone Mapping}


Despite being an ill-posed problem, automatic solutions are desired to ``up-convert'' existing SDR imagery such that it takes advantage of the extended capabilities of HDR displays and to generate alternative dynamic range versions of a single source image.
This is distinguished from the present application which aims to encode a transformation between SDR and HDR renderings that are already defined.

In recent years, deep-learning-based methods have been proposed to learn dynamic range conversion using HDR/SDR paired datasets by copying the analytical tone mapping approaches used in data generation. 
A number of approaches have been designed with model blocks specific to the SDR-to-HDR up-conversion task \cite{kim19,kim20,chen21}, but general image-to-image translation \cite{he16,isola17,zhu17} and photo editing methods \cite{gharbi17,he20,zeng20} can also be applied in this context.
While these approaches provide a promising solution to the problem of upconverting legacy content, they are not well suited for encoding since they are trained to reproduce a single tone mapping strategy and thus cannot account for a given user-specified solution.

In contrast, we propose a method for HDR reconstruction based on the gain map framework. Since the pixel-wise maps are not restrictive concerning the mapping strategy, this framework is amenable to the one-to-many nature of SDR-to-HDR mapping. 
Furthermore, the base-residual paradigm of the gain map framework provides an opportunity for novel encoding techniques which take advantage of the base image as a guide.
The first work to address gain maps in the imaging literature \cite{canham24}
shows that gain maps encoded with traditional compression can be improved with an alternative exponential residual. In the following sections, we demonstrate that these findings also apply to MLP-based approaches.

\subsection{Neural Implicit Functions}
The use of neural networks, in particular a multi-layer perceptron (MLP), to encode image data has been recently proposed \cite{sitzmann20,mildenhall21} as the neural functions described therein are continuous approximations of the input data, avoiding the need for traditional quantization. Many works ~\cite{strumpler2022implicit,dupont21,dupont22} have shown that neural implicit networks are effective for image compression. However, when used directly for compression, MLPs are still computationally expensive to optimize, as they rely only on positional coordinates $(x,y)$ to predict image values $(r,g,b)$. 
For this reason, current state of the art methods take on the order of minutes to days to encode a single image, video or 3-D scene \cite{lindell22,saragadam22,wu24}.
Our work is conceptually different as we aim to encode a spatially varying RGB transformation to be applied to an input SDR image, for which an INR can be trained in seconds.
As a result, our MLP is trained with position and corresponding SDR color values $(x,y,r,g,b)$ that are highly correlated with the gain map 
allowing it to be trained quickly. 

This setting was first proposed by Le et al.~\cite{le23}, who used an MLP with $(x,y,r,g,b)$ inputs to reconstruct clipped color values from gamut mapping. Liu et al.~\cite{liu24} later introduced a spatial $(x,y)$ and color $(r,g,b)$ network pair for embedded inverse tone mapping,  
incorporating online hard example mining (OHEM) and domain-wise pre-training.  
However, both networks directly output HDR RGB values.  We show in Sec.~\ref{sec:exp} that our approach, designed for an exponential residual encoding of the gain map,  
outperforms previous works in HDR reconstruction quality across multiple metrics.

\section{Proposed Method}
\label{methods}

\begin{figure*}[t]
\centering
\includegraphics[width=1\textwidth]{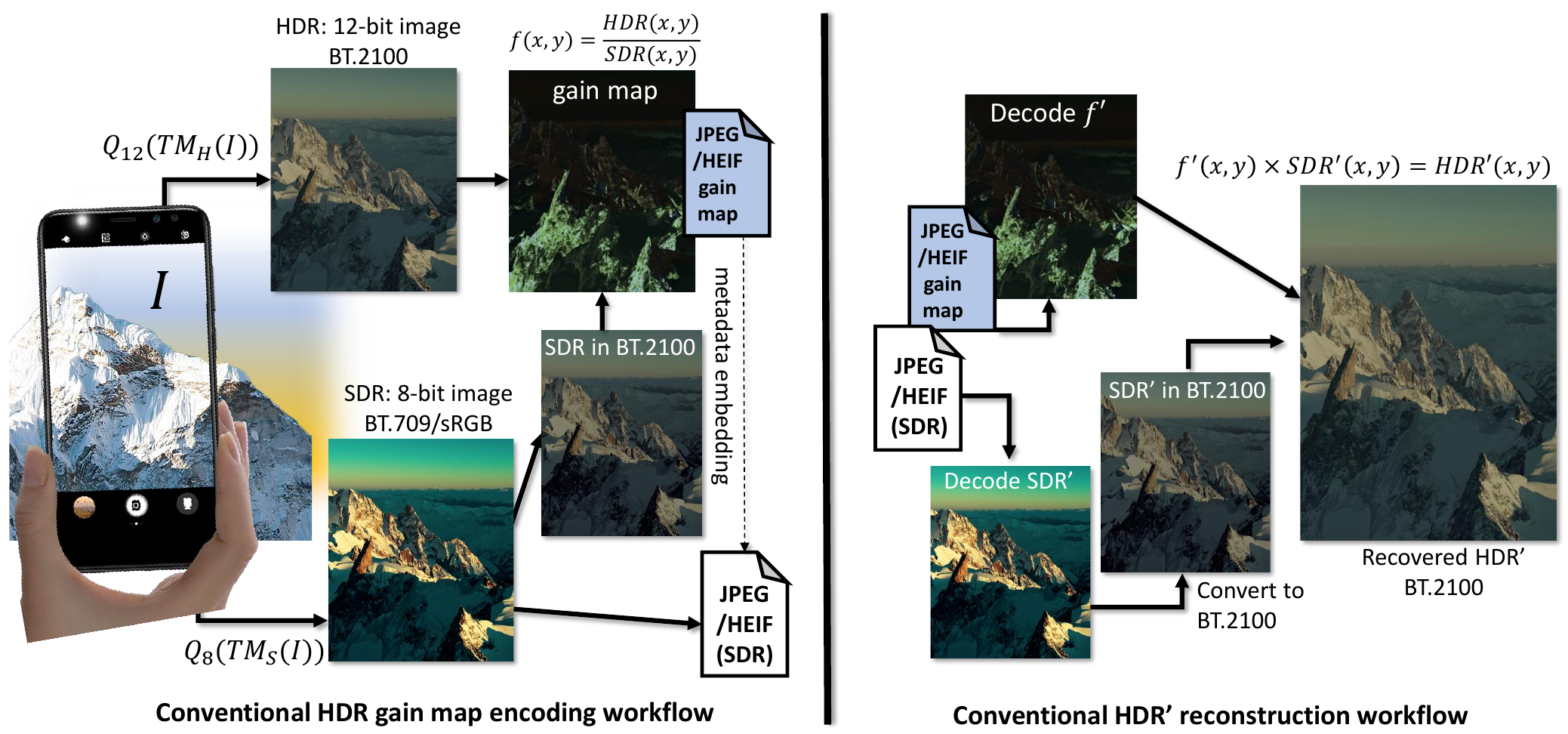}
\caption{Given an HDR scene with bright specular highlights, dark shadows, and smooth
gradients, the image is captured and mapped to two different renderings
intended for HDR and SDR display, respectively. Since most applications and displays
are designed around 8-bit SDR images and since they have a relatively smaller data
footprint than HDR images, it is advantageous to encode this version with JPEG, HEIC or JPEG-XL compression along
with metadata from which the HDR version can be reconstructed.}
\label{fig:overview}
\end{figure*}

The conventional framework recommended for HDR reference reconstruction given a lightweight metadata object is shown in Fig. \ref{fig:overview}. Given an HDR scene with specular highlights, shadows, and smooth gradations, sensors from smartphones to cinema cameras capture images in 10-12 bit representation that are mapped for standard dynamic range rendering, quantized to 8 bits (N-bit quantization is denoted as $Q_N(\cdot)$) and stored in a compressed file for general interchange. However, if an alternative HDR rendering is calculated before encoding, a pixel-wise gain map can summarize the ratio between the two alternative versions and be embedded in the SDR file to allow for reconstruction of the HDR version. The pixel-wise nature of the mapping allows for arbitrary re-mapping operations to be encoded, including global and local non-linear processing.

\subsection{Conventional Framework}

As described in the documents of Adobe \cite{adobe} and Android \cite{android}, a gain map is computed by taking the ratio of the HDR and SDR renderings. 
The framework begins with original images $S$ and $H$, encoded for a standard and a high dynamic range display respectively. First, the color encoding of the two sources is equalized by transforming the SDR image from its native space into the space of the HDR image as in Fig. \ref{fig:tm}. Then, the gain map is given by:
\begin{equation}
\label{eq:one}
f(x,y) = \frac{(H + \epsilon)}{(S + \epsilon)}, 
\end{equation}
where $\epsilon$ is a small offset applied to each image to avoid divide-by-zero errors. While it would be beneficial to calculate the gain map with a compressed and decoded version of $S$, we use a quantized, pre-encoding version for our experiments as our analysis focuses on residual encoding.

Next, $f(x,y)$ is normalized for JPEG-encoding (gain map values must be in range $[0,1]$, and bits should be allocated such that gradations are not lost in quantization).
This is accomplished by computing the bounds of the gain map, $f_\text{min}$ and $f_\text{max}$, and then applying a normalization of the gain map in a $log_2$ space as follows: 

\begin{equation}
\label{eq:two}
f_\text{norm}(x,y) = \frac{(log_2(f(x,y)) - log_2(f_\text{min}))}{(log_2(f_\text{max})-log_2(f_\text{min}))}.
\end{equation}
    
The output $f_{norm}(x,y)$ is clamped to range $[0,1]$, the gain map is then quantized to 8-bit precision, down-sampled, and compressed at a quality setting of no higher than 90 out of 100 \cite{android}. Using a multi-picture format (e.g. HEIF, AVIF, JPEG-XL) the map can be stored alongside $S$.
The parameters $log_2(f_\text{max})$, $log_2(f_\text{min})) $ and $\epsilon$ are then stored in metadata.
This strategy allows for $S$ to be interpreted by legacy applications while storing the necessary information to reconstruct $H$.

The encoding/decoding process returns a slightly different function $f'_\text{norm}(x,y)$ due to quantization and compression. 
The decoding process starts by converting back to a floating point representation in the range $[0,1]$ and the normalization is inverted as follows: 
\begin{equation}
\label{eq:three}
f'(x,y) = 2^{(f_\text{norm}(x,y)) * (log_2(f_\text{max})-log_2(f_\text{min})) + log_2(f_\text{min})}.
\end{equation}
Finally, the SDR image is decoded, converted to the HDR display space (as above) and the HDR rendition is reconstructed: 
\begin{equation}
\label{eq:four}
H' = (S + \epsilon) \odot f'(x,y) - \epsilon.   
\end{equation}

Canham et al.~\cite{canham24} show that gain maps encoded with traditional compression  can be improved by using an alternative residual, where Eqs.~\ref{eq:one} and \ref{eq:four}  
are replaced with Eqs.~\ref{eq:five} and \ref{eq:six}.  
This replaces a multiplicative residual (gain map) with an exponential residual (gamma map), as follows:
\begin{equation}
\label{eq:five}
f(x,y) = \frac{log(H + \epsilon)}{log(S + \epsilon)} 
\end{equation}
\begin{equation}
\label{eq:six}
H' = (S + \epsilon)^{f'(x,y)} - \epsilon.  
\end{equation}
Intuitively, an exponential residual acts as a more accurate predictive coding approximation of non-linear tone mapping operators than a multiplicative linear one. The experiments of Sec.~\ref{sec:exp} show that this approach significantly benefits the proposed method. 

\subsection{MLP Encoding Framework}

\begin{figure*}[t]
\centering
\includegraphics[width=1\textwidth]{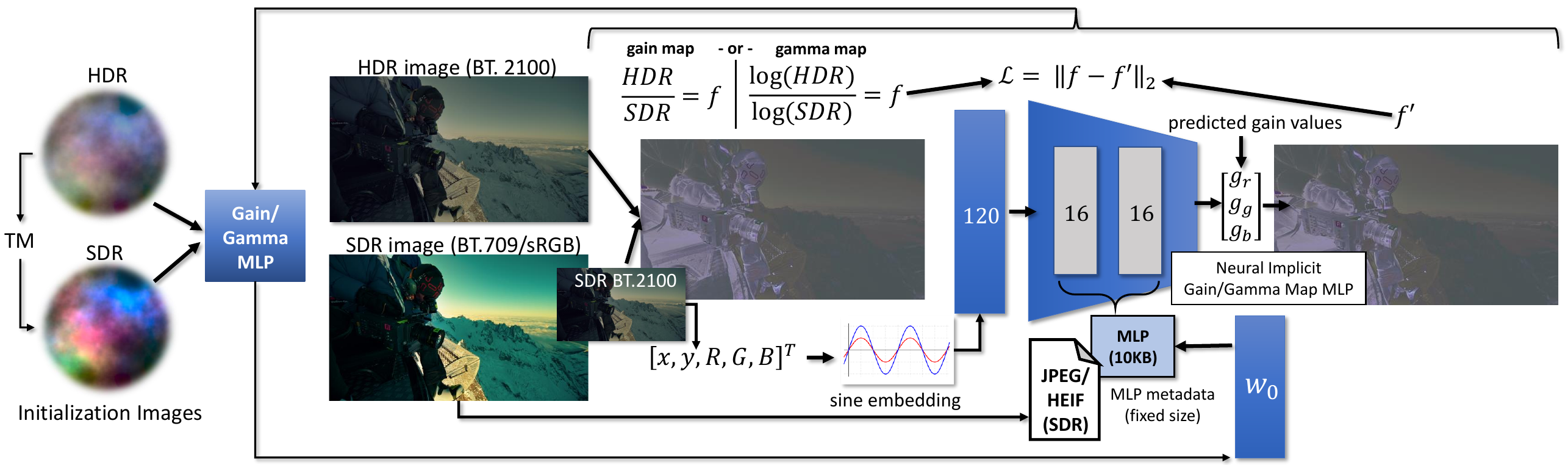}
\caption{The proposed MLP network architecture. Given the SDR image's pixel coordinates and RGB values, the network encodes each pixel in a 120-element sine embedding, which is then passed to a two-layer MLP, whose weights are optimized to predict the corresponding gain or gamma map values at each pixel. 
}
\label{fig:architecture}
\end{figure*}

The primary limitations of the conventional framework lie in the encoding of the gain map, $f(x,y)$, which is quantized to 8-bit precision, downsampled, and compressed. These steps introduce practical issues, as many tone mapping strategies apply tone scale adjustments that impact local details.  

As previously mentioned, our solution is based on neural implicit functions~\cite{sitzmann20,mildenhall21},  
which offer advantages due to their continuity and efficient image representation  
compared to traditional compression techniques used in conventional frameworks~\cite{adobe,android}.  Our network architecture is shown in Fig.~\ref{fig:architecture}.  
Given the $(x,y)$ coordinates and $R, G, B$ values of each pixel in the SDR image $S$,  
the network weights are optimized to return the corresponding ground truth pixel $f(x,y)$,  
which represents either the gain map or the exponential residual encoding, referred to as a gamma map.  
We denote the MLP's approximation of this function as $f'(x,y)$, as illustrated in Fig.~\ref{fig:architecture}.  

To optimize the MLP, we first append pixel coordinates to the RGB values to build a 5-D input $(x,y,r,g,b)$. Each of these inputs is converted to 24-D sinusoidal embedding, which results in a 120-D network input. The network is a two-layer ReLU MLP, resulting in a 10 KB model size. We use a hardware-aware fast-implementation to speed up optimization time~\cite{muller21}. Each layer contains 16 neurons, and the final layer returns the reconstructed gain per color channel $f'(x,y)$.  We use a batch size of $65,536$ randomly sampled locations within the image for each training iteration. Our loss is an MSE loss between the reconstructed gain map $f'(x,y)$ and ground truth gain map $f(x,y)$.  Finally, we use an Adam~\cite{adam} optimizer with a learning rate of 1e-2 and train for 1000 iterations.

Additionally, we employ a meta-initialization procedure from~\cite{meta, le23}  
to provide the MLP weights with a suitable initialization, accelerating optimization  
and improving reconstruction quality~\cite{meta}.    To compute these weights, the MLP model is optimized for 10,000 iterations on a set of 50 chromatic noise images,  
generated using spatio-chromatic natural image statistics through the process of Daly et al.~\cite{daly23},  
as shown in Fig.~\ref{fig:architecture}.  
The noise images span the Rec.~2020 gamut, while their BT.709 SDR counterparts are processed  
with the global, stylistically neutral default tone-mapping operator of DaVinci Resolve.  
We query the MLP at all pixel coordinates to reconstruct gain maps, which are then applied to the SDR image to reproduce the final HDR image.

\section{Experiments}
\label{sec:exp}
We demonstrate the effectiveness of our MLP-based approach on a variety of input images and tone mapping strategies by comparing state of the art HDR embedding networks (\cite{le23},\cite{liu24}) and traditional compression techniques (JPEG, HEIC, and JPEG-XL) with gain and gamma map residuals (Gain-MLP, Gamma-MLP). Additional contextual comparisons to the ITM and INR literature outside of the present encoding application are included in the supplementary material.

\subsection{Datasets}
\label{sec:data}
The tone mapping work of Cyriac et al. \cite{cyriac21} uses a set of manually graded SDR-HDR pairs to optimize their method. Starting from a series of 
HD (1920$\times$1080) SDR and HDR mastered images provided by Froehlich et al.~\cite{froehlich14} and the ARRI camera group \cite{arri} (encoded in BT.709 \cite{bt709} and BT.2020 \cite{bt2020}, respectively), a professional colorist produced an optimal visual match in the alternative format. 
A Sony BVM-X300 with a peak white luminance of 1,000 $cd/m^{2}$ and ST. 2084 \cite{st2084} decoding function was employed as a reference monitor and the colorist switched between HDR and SDR calibration profiles to compare the two renderings. This way, the manually graded pairs are derived from a modern HDR cinema workflow.

To account for different tone mapping operators, the  HDR images are processed by representative methods of the three major categories described by Ou et al. \cite{ou22}. Since these mapping functions are intended to deal with linear radiance values, the encoding function of the HDR images was inverted to return display radiance values and processed by the methods of Drago et al. \cite{drago03} (log-based), Reinhard et al. \cite{reinhard05} (exposure-based) and Larson et al. \cite{larson97} (histogram equalization) using the default parameters of the MATLAB HDR toolbox ~\cite{Banterle:2017}.

We also utilize the dataset of Chen et al.~\cite{chen21}, which has become a de facto standard for SDR-HDR mapping. This dataset consists of HDR10-encoded images (BT.2020) sourced from various movies and product test reels.  The corresponding SDR versions were generated using YouTube's automatic HDR-to-SDR processing,  resulting in 117 8-bit sRGB UHD (3840$\times$2160) images.  

\subsection{Comparisons}
\label{sec:comp}
The conventional gain map framework was implemented using the Python Image Library (PIL) as described above. Bicubic resizing was employed to obtain the $1/4$ resolution version and JPEG, HEIC and JPEG-XL gain map images were stored using the ``quality'' parameter of 80 out of 100.

The proposed method is tested in its default configuration using two 16-neuron layers. The optimization is run for 1000 iterations, resulting in a 4-second optimization time per image measured on an Nvidia Quadro RTX 6000. Method performance as a function of optimization time is demonstrated in the supplementary material. 


In Tab. \ref{table:ourData}, the datasets of Chen et al. \cite{chen21} (117 images $\times$ 1 tone mapping method) and Cyriac et al. \cite{cyriac21} (29 images $\times$ 4 tone mapping methods) were then pooled together to compare the proposed MLP approaches (Gain-MLP, Gamut-MLP) to the traditional compression techniques (JPEG, HEIC and JPEG-XL) for encoding gain and gamma map residuals.
Also, the model of Le et al. \cite{le23} (Direct-MLP) is reinitialized for comparison in the present application using the proposed synthetic initialization images.
Finally, the method of Liu et al. \cite{liu24} (MLP-ITM) is tested in its default setting (34 KB network, 2 second optimization time), using its domain-wise initialization.

The rate-distortion performance of these methods is also compared for the manually tone mapped data of Cyriac et al. \cite{cyriac21} in Fig. \ref{fig:rdPlot}. The associated documents \cite{adobe,android} recommend modulating the conventional approaches' bit rates by adjusting the resize factor of the gain maps. As such, JPEG, JPEG-XL and HEIC residuals were resized to $1/8, 1/4, 1/2$ and full resolutions. For the MLP based approaches (Gamma-MLP, Gain-MLP, Direct-MLP \cite{le23} and MLP-ITM \cite{liu24}), the number of nodes on each network layer was tested at levels 8, 16, 64 and 128 with a 4 second optimization time. To accomplish this, the meta-initializations were retrained at each model size. Each model was also reinitialized with its default parameters to confirm the consistency of this process.

Since our reconstruction is an information fidelity task, Peak Signal-to-Noise Ratio (PSNR) was used. However, since it is measured pixel-wise, it is not sensitive to structured compression and quantization artifacts. We also evaluate the Structural Similarity Index Metric (SSIM) \cite{wang04}. For both metrics, reconstructed and ground truth HDR images are passed directly in their native encoding. SSIM was designed for use with SDR images, so we also employ an HDR native metric (HDR-VDP3 \cite{mantiuk23}), which accounts for visual sensitivity to different luminance regimes. Finally, we test the well-known color difference metric $\Delta E_{00}$ and an HDR alternative, $\Delta E_{IPT}$, averaging pixel-wise differences over the image plane in both cases.

\subsection{Quantitative Results}
Tab. \ref{table:ourData} shows the comparison of the proposed MLP approach versus the standard framework with JPEG, HEIC and JPEG-XL in the context of gain map (Gain-MLP) and gamma map (Gamma-MLP) encoding averaged over all source images and tone mapped versions described in Sec. \ref{sec:data}.
These results demonstrate that implicit neural representation based on the proposed lightweight MLP can outperform traditional compression methods in a base-residual HDR encoding. Moreover, our MLP approach outperforms--in all but one metric--the prior state-of-the-art \cite{liu24} while only requiring one third of \cite{liu24} size.   While MLP-ITM~\cite{liu24} achieves the best overall performance for the HDR-VDP3 metric, our method performs on par with the second-best results.  

\begin{figure}[]
\centering
\includegraphics[width=0.5\textwidth]{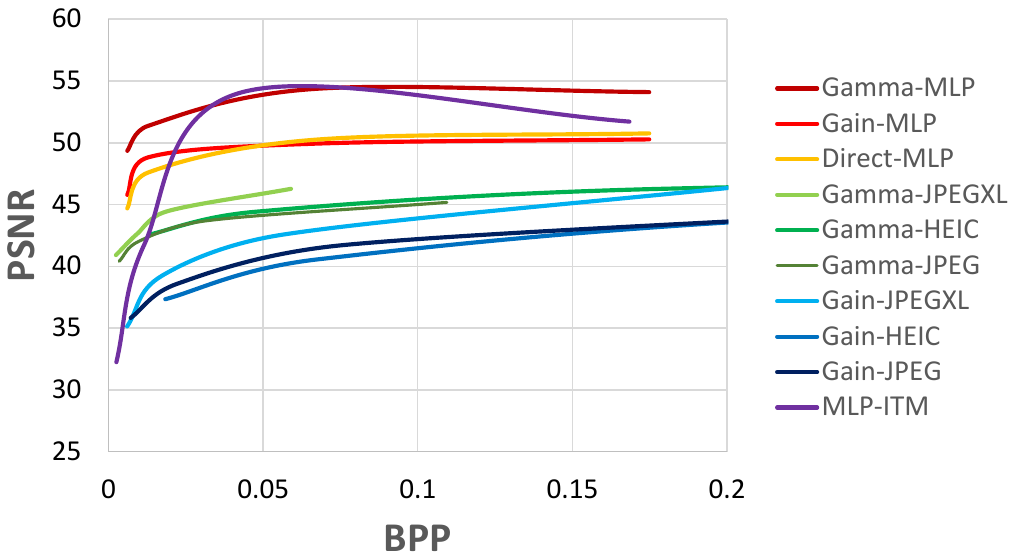}
\caption{Rate-distortion plot for manually tone mapped version of Cyriac et al \cite{cyriac21}. The down-sampling parameter of traditional compression methods is adjusted to levels $1/8, 1/4, 1/2$, and full resolution. MLP based methods are adjusted to 8, 16, 64, and 128 nodes and are set to equivalent optimization times. 
The resulting compressed residual sizes are divided by the image resolution $1920 \times 1080 \times 3$ to obtain bits-per-pixel (BPP) values.}\vspace{-6mm}
\label{fig:rdPlot}
\end{figure}

Fig. \ref{fig:rdPlot} demonstrates that exponential gamma map residuals consistently allow for high quality HDR reconstructions even at very low bit rates in traditional compression and MLP-based encoding paradigms.

\begin{table*}[t]
\begin{center}
\begin{tabular}{lccccccc} \hline
\textbf{Method} & PSNR $\uparrow$ & $\Delta E_{00}$ $\downarrow$ & SSIM $\uparrow$ & $\Delta E_{IPT}$ $\downarrow$ & HDR-VDP3 \cite{mantiuk23} $\uparrow$ & Size (KB) $\downarrow$ \\ \hline
Gain-JPEG & 38.29 & \textcolor{red}{\textbf{2.16}} & 0.968 &  \textcolor{red}{\textbf{9.63}} & 7.92 & 19.0 \\
Gamma-JPEG & 41.45 & 1.37 & 0.979 &  7.12 & 8.62 & 19.4 \\
Gain-HEIC & 39.20 & 1.98 & 0.972 &  8.71 & 8.14 & 18.4 \\
Gamma-HEIC & 42.21 & 1.27 & 0.982 &  6.57 & 8.75 & 18.2 \\
Gain-JPEG-XL & \textcolor{red}{\textbf{37.65}} & 1.73 & \textcolor{red}{\textbf{0.950}} & 8.21 & \textcolor{red}{\textbf{7.82}} & 11.4 & \\
Gamma-JPEG-XL & 40.58 & 1.15 & 0.957 & 6.40 & 8.35 & 12 \\
Direct-MLP \cite{le23} & 46.30  & 0.96 & 0.988 & 4.66 & 9.06 & 10 \\
MLP-ITM \cite{liu24} & 47.25 & 0.87 & 0.991 &  4.28 & \textbf{9.13} & \textcolor{red}{\textbf{34}} \\
Gain-MLP (ours) & 47.60 & 1.02 & 0.992 & 4.27 & 8.98 & \textbf{10} \\ 
Gamma-MLP (ours) & \textbf{48.53} & \textbf{0.78} & \textbf{0.993} & \textbf{3.91} & 9.11 & \textbf{10} & \\


\end{tabular}

\caption{Quantitative comparison between traditional encoding techniques, the state-of-the-art application relevant INR techniques \cite{le23, liu24} and the proposed MLP averaged over Cyriac et al. \cite{cyriac21}, and Chen et al. \cite{chen21} datasets. In the gain map configuration, a multiplicative residual between the HDR and SDR images is encoded, while an exponential residual is encoded in the gamma map configuration. The best-performing method is \textbf{bold}, while the worst performing method is in \textcolor{red}{\textbf{red}}.}
\label{table:ourData}
\end{center}

\end{table*}

\subsection{Qualitative Results}
\begin{figure*}
\centering
\includegraphics[width=1.0\textwidth]{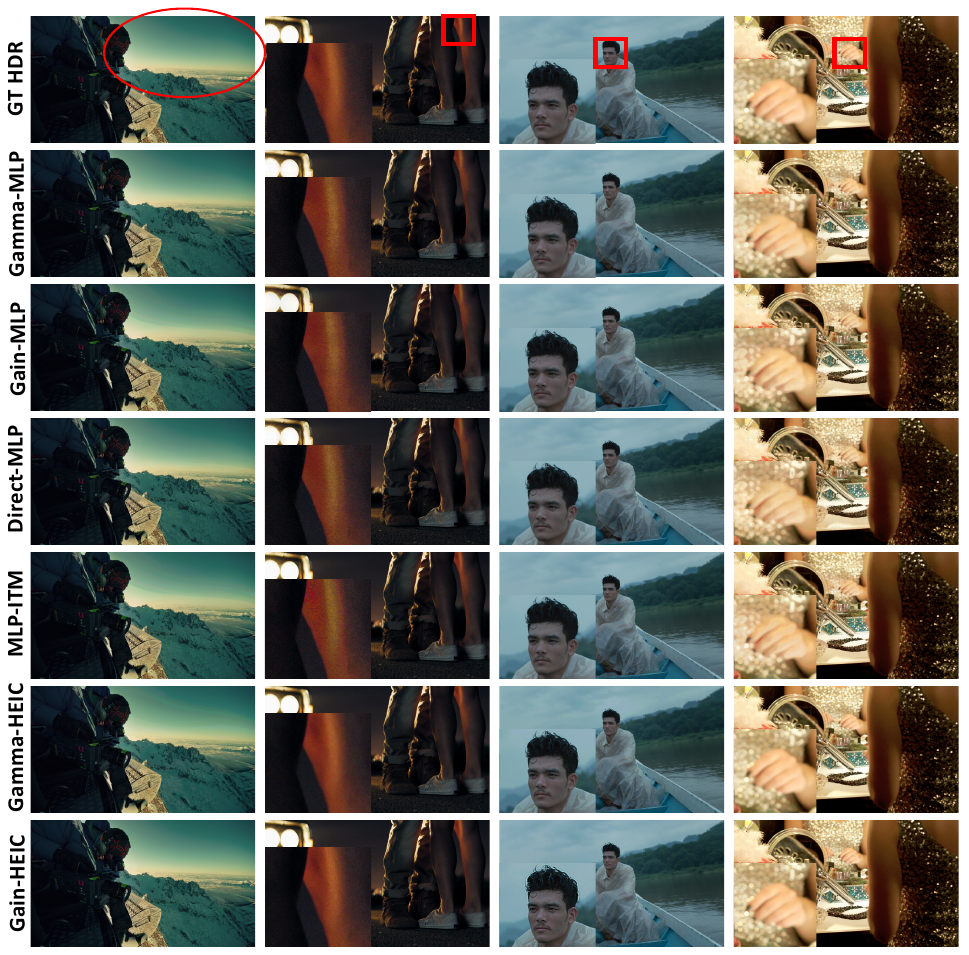}
\caption{Qualitative comparison between the best performing methods of the quantitative comparison on images of the dataset of Cyriac et al. \cite{cyriac21}. Here, HDR reconstructions from manually tone mapped SDR images are transformed to an intermediate display space for visualization (DCI-P3, 2.4 gamma). Traditional compression results show banding, haloing, block based artifacts and loss of high frequency detail. Color errors, flattened details and banding can be observed with the MLP-based methods, but artifacts are reduced in both MLP and traditional compression cases by the use of gamma maps.}
\label{fig:qualitative1}
\end{figure*}

Fig. \ref{fig:qualitative1} shows comparisons between the proposed and conventional HDR encoding approach for several images from the Cyriac et al. \cite{cyriac21} source dataset with the tone mapping of Reinhard et al. \cite{reinhard05} applied.
HDR reconstructions are transformed from BT.2020 to P3-D65, a medium gamut color space that many consumer devices already achieve, to better simulate their HDR-displayed appearance.
The results show that traditional compression methods suffer from banding, haloing, block based artifacts, and loss of high frequency detail. To a lesser degree, the MLP-based methods also suffer from banding, color errors and loss of clipped details, but artifacts are visibly reduced by the use of the exponential residual encoding (gamma map) for both the proposed MLP approach and traditional methods.

\section{Discussion \& Limitations}

In traditional compression, images are divided into blocks, transformed into the frequency domain, and their high-frequency components are quantized. This prioritizes bit rate allocation to low-frequency regions, which are more common in natural images and where the human visual system is more sensitive to quantization artifacts.  In addition, such frequency-quantization is often applied uniformly regardless of image content.  
Additionally, they have no built-in mechanism to take advantage of the available SDR image that is readily available as part of the encoding framework.  In our case, the MLP-based framework is directly optimized to consider the SDR image and the target output, which are highly correlated (even when residuals are not used, as in Direct-MLP \cite{le23} and MLP-ITM \cite{liu24}).

Fig. \ref{fig:rdPlot} demonstrates that the base-residual approach of gain maps benefits MLP-based encoding approaches, as Gain-MLP outperforms Direct-MLP at low bit rates.
Furthermore, we show that using exponential residuals allows for even higher quality and more efficient representations of tone mapping transforms, allowing for state-of-the-art methods to be outperformed by a simpler network (Gamma-MLP) which is more stable across the range of tested bit rates. 
Intuitively, this could be explained by the fact that the residual computation process more closely resembles tone mapping operations acting as a predictive coding step. 

In the supplementary material, the results of Tab. \ref{table:ourData} are broken down among all tone mapped versions of the two source datasets.
While the method rankings are rarely upset, there is an overall difference in scaling the metric scores when using different SDR tone mapped versions.
The main relevant difference between these tone mappers is their degree of clipped information, which the MLPs have a limited ability to reconstruct within practical optimization times (as demonstrated in Fig. \ref{fig:qualitative1} and the supplementary material).

\section{Concluding Remarks}

This paper proposes an MLP-based framework alternative for gain map compression. 
Our approach takes an SDR image as input and optimizes a lightweight multi-layer perceptron network to predict its corresponding SDR-to-HDR residual.  The resulting MLP weights can be stored in less than 10 KB of metadata.  We demonstrate that a network of this size allows for reconstructions that exceed the quality of traditional compression. We also showed that swapping the multiplicative residual for an exponential one results in a robust model which outperforms state-of-the-art embedded inverse tone mapping approaches at a range of bit rates.
Also, we extend a pre-existing manual tone mapping dataset with additional SDR images from canonical tone mappers to more thoroughly characterize the problem space and we demonstrate the novel use of chromatic noise patterns reflecting natural image statistics as initialization data for INR models.
As future work we propose a mechanism to adjust MLP encoding bit rate to allow for consistent reconstruction performance for images of varying resolutions.
Code and images used in this paper will be made available upon acceptance.

{
    \small
    \bibliographystyle{ieeenat_fullname}
    \bibliography{main}
}
\clearpage
\setcounter{page}{1}
\maketitlesupplementary







\maketitle

This supplementary material provides (1) ablation experiments to characterize the performance of our proposed method and related works in terms of reconstruction quality; (2) additional figures demonstrating the different tone mapping methods employed in the dataset; (3) an expanded results table demonstrating method performance on individual tone mapping strategies.


\section{Ablations} \label{ablations}
\begin{figure} [h!]
\centering
\includegraphics[width=0.5\textwidth]{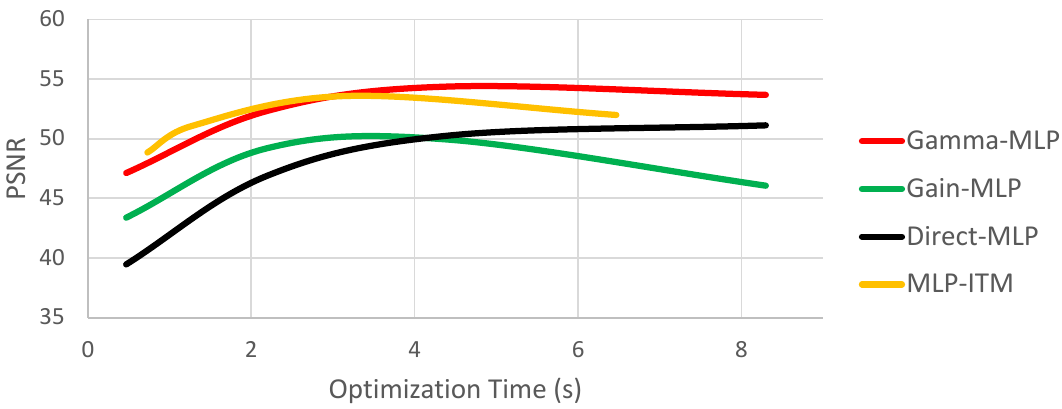}
\caption{MLP performance as a function of optimization time on the manually tone-mapped data from the Cyriac et al. dataset \cite{cyriac21}. All networks were set to match the metadata size of MLP-ITM in its default setting \cite{liu24}. Gain-MLP has higher performance than Direct-MLP at low optimization times, but Direct-MLP improves across tested optimization times. Similarly, MLP-ITM has best performance at low optimization times but Gamma-MLP has an extended lattitude for improvement.} 
\label{fig:time_plot}
\end{figure}

In addition to network size, which is examined in the main text, MLP performance is dependent on optimization time. In Fig. \ref{fig:time_plot}, we test the MLP methods addressed in the main paper on the manually tone mapped images of Cyriac et al. \cite{cyriac21} at different optimization times. In these experiments, network size is matched to the default settings of the embedded ITM baseline, MLP-ITM, since its performance sharply decreases outside of this setting. The results show that this network is roughly as computationally efficient as the proposed Gamma-MLP. When compared to Direct-MLP \cite{le23}, Gain-MLP has higher performance at low optimization times but Direct-MLP sees consistent improvements with increased optimization time.

Next, we compare Gain-MLP performance with and without the SDR base image as input. As mentioned in the main paper, applying MLPs directly for compression~\cite{strumpler2022implicit,dupont21,dupont22} can still be computationally expensive to optimize, as this approach relies only on positional coordinates $(x,y)$ to predict image values $(r,g,b)$. Our work has a significant advantage, given our access to the SDR image to guide the reconstruction of the gain map. Table \ref{table:positionAblation} shows how Gamma-MLP's performance reduces significantly if the input is only positional coordinates.

\begin{table}
\begin{center}

\begin{tabular}{lccccc} 
 \toprule  
Model Inputs & PSNR $\uparrow$ & SSIM $\uparrow$ & HDR-VDP3 $\downarrow$ \\
\hline
\noalign{\smallskip}
 (x,y) & 36.6 & 0.969 & 7.95 \\
 (x,y,r,g,b) & 51.4 & 0.998 & 9.74 \\
 \hline
\hline 
\end{tabular}
\caption{Ablation of MLP (2-layer, 16 nodes) performance when predicting gamma map with just position information (x,y) compared to having access to position information (x,y) and the SDR RGB (r,g,b) value for prediction. We report results on the manually tone-mapped dataset. The experiment demonstrates the benefit of the SDR guide image in our application.}
\label{table:positionAblation}
\end{center}

\end{table}
\setlength{\tabcolsep}{1.4pt}

Finally, in Table \ref{table:deepnets} the Gamma-MLP results are compared against the deep learning based inverse tone mapping and image translation methods tested by Chen et al.~\cite{chen21} on their 117-image UHD test set. 
We show that when trained on a particular tone mapper, these methods are outperformed in the reproduction of that function by standard Gain map encoding using HEIC, so they are not a viable replacement for gain map encoding.

\begin{table}[h]
\begin{center}

\centering
\begin{tabular}{lccc}
    \toprule    
    \textbf{Method} & PSNR $\uparrow$ & SSIM $\uparrow$ & HDR-VDP-3 $\uparrow$ \\
    \hline
    \noalign{\smallskip}
    ResNet \cite{he16} & 37.3 & 0.972 & 8.39 \\
    Pix2pix \cite{isola17} & 25.8 & 0.878 & 7.14 \\
    CycleGAN \cite{zhu17} & 21.3 & 0.850 & 6.94 \\
    HDRNet \cite{gharbi17} & 35.7 & 0.966 & 8.45 \\
    CSRNet \cite{he20} & 35.0 & 0.963 & 0.14 \\
    Ada-3DLUT \cite{zeng20} & 36.2 & 0.966 & 2.37 \\
    Deep SR-ITM \cite{kim19} & 37.1 & 0.969 & 11.48 \\ 
    JSI-GAN \cite{kim20} & 37.0 & 0.969 & 4.24 \\
    AGCM+LE \cite{chen21} & 37.6 & 0.973 &	5.64 \\
    AGCM \cite{chen21} & 36.9 & 0.966 & 0.280 \\
    Gain-HEIC & 39.2 & 0.972 & 0.02 \\
    ITM-MLP \cite{liu24} & \textbf{41.6} & \textbf{0.988} & 0.04 \\
    Gain-MLP & 41.2 & 0.986 & \textbf{0.01} \\ 
    Gamma-MLP & 41.5 & \textbf{0.988} & \textbf{0.01} \\\hline
    \end{tabular}
  
\end{center}

\caption{Using the 117 UHD image test dataset of Chen et al. \cite{chen21}, where HDR graded masters are automatically converted to SDR using Youtube's HDR10 pipeline, the proposed MLP (Gain-MLP) is compared to deep learning methods for HDR generation (optimized on the associated training set). While the application of deep-learning methods differs from ours (automatic upconversion of SDR content with no prior knowledge of an HDR reference), we outperform their reconstructions by encoding transforms on an image-to-image basis with a lightweight model. Best results highlighted in green.}
  \label{table:deepnets}
\end{table}
\setlength{\tabcolsep}{1.4pt}

\section{Experiments} \label{Experiments}

\begin{figure*} [h]
\centering
\includegraphics[width=1\textwidth]{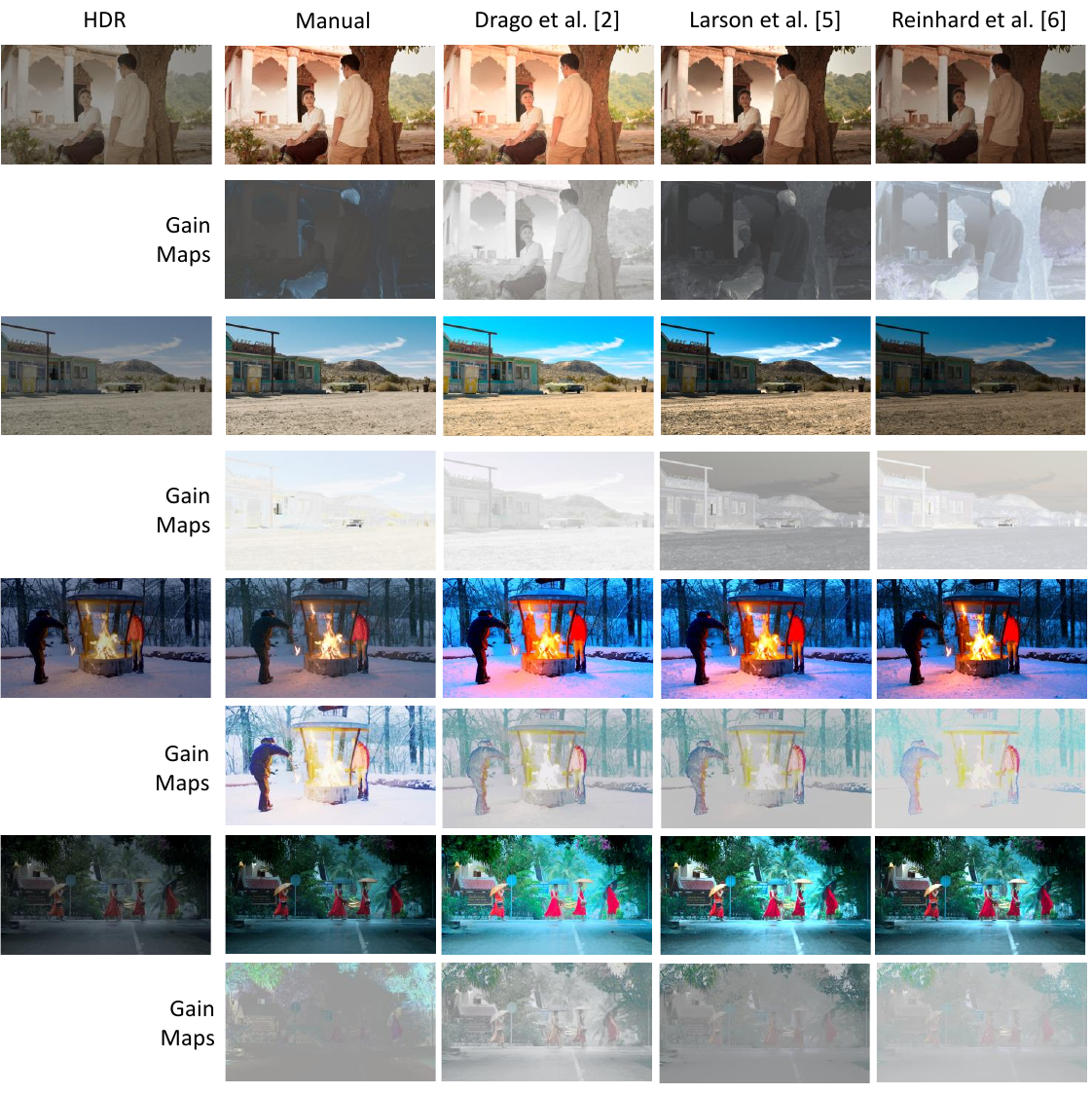}
\caption{The tone mapping methods tested in the experimental section \cite{drago03,larson97,reinhard05} result in qualitatively different SDR renditions of the HDR source. The methods' gain maps are visualized to illustrate how the HDR reconstruction task varies for different tone mapping conditions. In some instances, the results are brighter or darker, sharper, or more saturated. In general, image detail is preserved better in manual tone mapping, while automatic tone mapping tends to over-saturate dark images and amplify noise (e.g., the last two rows.)} 
\label{fig:tmMethods}
\end{figure*}

In the experimental section of the main paper, the dataset of Cyriac et al. \cite{cyriac21} is employed to evaluate the performance of the proposed method (Gamma-MLP) against the existing framework. We augment this dataset for our quantitative comparisons by processing the HDR image through additional automatic tone mapping methods to complement the manual tone mapping results. In Fig. \ref{fig:tmMethods}, the performance of the different tone mapping methods is evaluated qualitatively. The gain maps demonstrate that these represent significantly different reconstruction tasks, explaining the variation in their respective results. In addition, it is demonstrated that the automatically tone-mapped images represent challenging cases as they are more likely to saturate detail and amplify noise.

While the results in the main paper were averaged over all tone mapping methods from the dataset of Cyriac et al. and Chen et al. \cite{chen21}, Table \ref{table:quantExt} splits the results between these subsets. It can be observed that in the conventional approach, reconstruction quality varies significantly between tone mapping methods for the Cyriac dataset, but the proposed MLP is relatively consistent between these subsets. Comparing results from different HDR sources (between Cyriac et al. and Chen et al. datasets) the conventional approach maintains reconstruction quality but its size increases, while the proposed MLP has a lower quality reconstruction if the network size is maintained.

\begin{table*}[ht!]
\begin{center}

\begin{tabular}{cccccccc} 
 \toprule  
& & PSNR $\uparrow$ & $\Delta E_{00}$ $\downarrow$ & SSIM $\uparrow$ & $\Delta E_{IPT}$ $\downarrow$ & HDR-VDP3 $\uparrow$ & Size (KB) $\downarrow$ \\ 
\hline

\noalign{\smallskip}

        ~ & Gain-JPEG & 38.1 & \textbf{\textcolor{red}{1.81}} & \textbf{\textcolor{red}{0.9721}} & \textbf{\textcolor{red}{8.86}} & 8.13 & 15  \\ 
        ~ & Gain-HEIC & 39.0 & 1.63 & 0.9768 & 7.93 & 8.42 & 14  \\ 
        ~ & Gain-JPEG-XL & \textbf{\textcolor{red}{37.2}} & 1.44 & 0.9435 & 7.71 & \textbf{\textcolor{red}{7.94}} & 10  \\ 
        ~ & Gamma-JPEG & 44.3 & 0.89 & 0.9874 & 5.17 & 9.26 & 14  \\ 
        Cyriac et al. \cite{cyriac21} & Gamma-HEIC & 45.2 & 0.80 & 0.9891 & 4.68 & 9.37 & 12  \\ 
        Manual TM & Gamma-JPEG-XL & 43.1 & 0.76 & 0.9553 & 4.78 & 8.96 & \textbf{9}  \\ 
        ~ & MLP-iTM & 51.1 & \textbf{0.53} & 0.9974 & \textbf{2.32} & \textbf{9.83} & \textbf{\textcolor{red}{34}}  \\ 
        ~ & Direct-MLP & 47.6 & 0.92 & 0.9918 & 3.61 & 9.50 & 10  \\ 
        ~ & Gain-MLP & 48.8 & 0.99 & 0.9945 & 3.59 & 9.15 & 10  \\ 
        ~ & Gamma-MLP & \textbf{51.4} & \textbf{0.53} & \textbf{0.9978} & \textbf{2.32} & 9.74 & 10 \\ 
        \hline

        ~ & Gain-JPEG & 39.0 & \textbf{\textcolor{red}{2.08}} & 0.9691 & \textbf{\textcolor{red}{9.08}} & 8.01 & 15  \\ 
        ~ & Gain-HEIC & 39.9 & 1.92 & 0.9731 & 8.22 & 8.24 & 13  \\ 
        ~ & Gain-JPEG-XL & \textbf{\textcolor{red}{38.2}} & 1.55 & \textbf{\textcolor{red}{0.9482}} & 7.23 & \textbf{\textcolor{red}{7.99}} & \textbf{9}  \\ 
        ~ & Gamma-JPEG & 42.5 & 1.23 & 0.9814 & 6.28 & 8.85 & 15  \\ 
        Cyriac et al. \cite{cyriac21}  & Gamma-HEIC & 43.3 & 1.13 & 0.9837 & 5.72 & 8.98 & 13  \\ 
        Reinhard et al. \cite{reinhard05} & Gamma-JPEG-XL & 41.4 & 0.97 & 0.9550 & 5.33 & 8.66 & 11  \\ 
        ~ & MLP-iTM & 48.8 & 0.86 & 0.9908 & 4.09 & 9.22 & \textbf{\textcolor{red}{34}}  \\ 
        ~ & Direct-MLP & 47.7 & 0.91 & 0.9883 & 4.48 & 9.18 & 10  \\ 
        ~ & Gain-MLP & 49.8 & 0.79 & \textbf{0.9933} & 3.47 & 9.25 & 10  \\ 
        ~ & Gamma-MLP & \textbf{50.6} & \textbf{0.69} & \textbf{0.9933} & \textbf{3.34} & \textbf{9.29} & 10 \\ 
        
        \hline

        ~ & Gain-JPEG & 36.0 & \textbf{\textcolor{red}{2.43}} & 0.9510 & \textbf{\textcolor{red}{11.23}} & \textbf{\textcolor{red}{7.33}} & 16  \\ 
        ~ & Gain-HEIC & 36.8 & 2.29 & 0.9565 & 10.33 & 7.60 & 14  \\ 
        ~ & Gain-JPEG-XL & \textbf{\textcolor{red}{35.2}} & 1.87 & \textbf{\textcolor{red}{0.9305}} & 9.31 & 7.34 & 10  \\ 
        ~ & Gamma-JPEG & 41.0 & 1.28 & 0.9753 & 6.75 & 8.56 & 16  \\ 
        Cyriac et al. \cite{cyriac21} & Gamma-HEIC & 41.6 & 1.20 & 0.9776 & 6.26 & 8.71 & 15  \\ 
        Larson et al. \cite{larson97} & Gamma-JPEG-XL & 39.9 & 1.01 & 0.9491 & 5.78 & 8.39 & \textbf{9}  \\ 
        ~ & MLP-iTM & 48.3 & \textbf{0.70} & \textbf{0.9948} & \textbf{3.47} & \textbf{9.39} & \textbf{\textcolor{red}{34}}  \\ 
        ~ & Direct-MLP & 47.6 & 0.76 & 0.9912 & 3.77 & 9.32 & 10  \\ 
        ~ & Gain-MLP & 48.0 & 0.75 & 0.9955 & 3.40 & 9.18 & 10  \\ 
        ~ & Gamma-MLP & \textbf{49.6} & 0.73 & 0.9930 & 3.84 & 9.14 & 10 \\ 
        
        \hline

        ~ & Gain-JPEG & 41.2 & \textbf{\textcolor{red}{1.89}} & 0.9731 & \textbf{\textcolor{red}{8.24}} & 8.54 & 12  \\ 
        ~ & Gain-HEIC & 42.3 & 1.70 & 0.9769 & 7.34 & 8.69 & 10  \\ 
        ~ & Gain-JPEG-XL & \textbf{\textcolor{red}{40.6}} & 1.39 & \textbf{\textcolor{red}{0.9509}} & 6.47 & \textbf{\textcolor{red}{8.38}} & \textbf{7}  \\ 
        ~ & Gamma-JPEG & 42.6 & 1.33 & 0.9780 & 6.66 & 8.80 & 13  \\ 
        Cyriac et al. \cite{cyriac21} & Gamma-HEIC & 43.3 & 1.24 & 0.9803 & 6.15 & 8.91 & 11  \\ 
        Drago et al. \cite{drago03} & Gamma-JPEG-XL & 41.7 & 1.06 & 0.9513 & 5.67 & 8.63 & 9  \\ 
        ~ & MLP-iTM & 46.5 & 1.05 & 0.9881 & 5.48 & 9.01 & \textbf{\textcolor{red}{34}}  \\ 
        ~ & Direct-MLP & 47.6 & 0.86 & 0.9861 & 4.75 & 9.06 & 10  \\ 
        ~ & Gain-MLP & \textbf{49.6} & 0.74 & 0.9920 & 3.99 & 9.11 & 10  \\ 
        ~ & Gamma-MLP & \textbf{49.6} & \textbf{0.73} & \textbf{0.9930} & \textbf{3.84} & \textbf{9.14} & 10 \\ 
        
        \hline

        ~ & Gain-JPEG & 37.0 & \textbf{\textcolor{red}{2.58}} & \textbf{\textcolor{red}{0.9743}} & 10.73 & 7.57 & 37  \\ 
        ~ & Gain-HEIC & 38.0 & 2.35 & 0.9780 & 9.75 & 7.76 & 41  \\ 
        ~ & Gain-JPEG-XL & 37.1 & 2.40 & 0.9757 & 10.32 & \textbf{\textcolor{red}{7.45}} & 21  \\ 
        ~ & Gamma-JPEG & 36.9 & 2.10 & 0.9747 & \textbf{\textcolor{red}{10.74}} & 7.66 & 39  \\ 
        Chen et al. \cite{chen21} & Gamma-HEIC & 37.7 & 2.03 & 0.9774 & 10.05 & 7.77 & \textbf{\textcolor{red}{40}}  \\ 
        Youtube TM  & Gamma-JPEG-XL & \textbf{\textcolor{red}{36.8}} & 1.95 & 0.9729 & 10.47 & 7.15 & 22  \\ 
        ~ & MLP-iTM & \textbf{41.6} & \textbf{1.22} & \textbf{0.9879} & \textbf{6.04} & 8.25 & 34  \\ 
        ~ & Direct-MLP & 41.1 & 1.35 & 0.9861 & 6.67 & 8.22 & 10  \\ 
        ~ & Gain-MLP & 41.2 & 1.56 & 0.9861 & 6.88 & 8.19 & 10  \\ 
        ~ & Gamma-MLP & 41.5 & 1.25 & 0.9876 & 6.19 & \textbf{8.26} & \textbf{10} \\ 
        
        \hline

\end{tabular}
\caption{Quantitative comparison between traditional encoding techniques and the proposed MLP broken up for tone mapping (TM) variations of Cyriac et al. \cite{cyriac21} (Manual, Drago et al. \cite{drago03}, Larson et al. \cite{larson97}, Reinhard et al. \cite{reinhard05}), and Chen et al. \cite{chen21} (Youtube)  datasets. The Direct-MLP strategy directly predicts the HDR image given the SDR input. In the gain map configuration, a multiplicative residual between the HDR and SDR images is encoded instead.
The results show that tone mapping variations on the same dataset make less of an impact on reconstruction quality for Gain-MLP than with the conventional approach. Best results per subset highlighted in green.}
\label{table:quantExt}
\end{center}
\end{table*}
\setlength{\tabcolsep}{1.4pt}

\begin{figure} [t]
\centering
\includegraphics[width=0.5\textwidth]{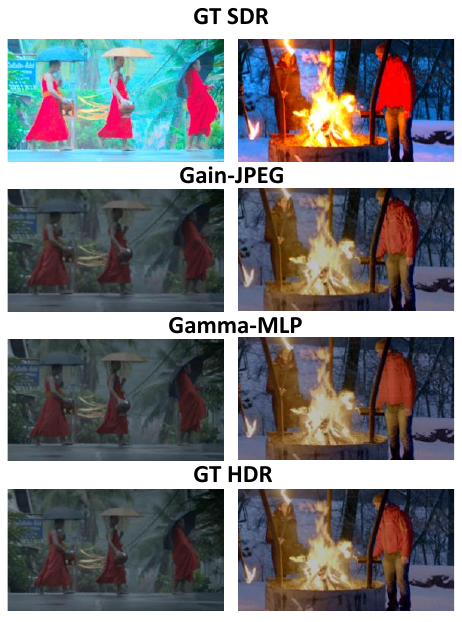}
\caption{The proposed method (Gamma-MLP) produces artifacts when the ground truth SDR image has lost significant detail due to over-saturation in the tone mapping stage. As a result, detail is lost (e.g., the monk's robes in the top row and the red jacket in the bottom row). The conventional framework results also suffer in these scenarios (Gain-JPEG included for comparison).} 
\label{fig:artifacts}
\end{figure}

Next, representative examples of the artifacts that can occur with the proposed MLP are demonstrated in Fig. \ref{fig:artifacts}. When large areas of detail are lost in the SDR image during the tone mapping process, more information is required to reconstruct the area, pushing the representation limits of the proposed and conventional methods.


\end{document}